\documentstyle[epsf,prd,aps,floats]{revtex} 
\newcommand{\be}{\begin{equation}}   
\newcommand{\ee}{\end{equation}}   
\newcommand{\bea}{\begin{eqnarray}}   
\newcommand{\eea}{\end{eqnarray}}

\begin{document}    
\twocolumn[\hsize\textwidth\columnwidth\hsize\csname@twocolumnfalse\endcsname    
\author{Stephon Alexander$^{1)}$, Robert Brandenberger$^{2)}$ and
 Jo\~ao Magueijo$^{1)}$}   
\date{\today} 
\address{1) Theoretical Physics, The Blackett Laboratory, 
Imperial College, Prince Consort Road, London, SW7 2BZ, U.K.}
\address{2) Physics Department, Brown University, Providence,
RI 02912, USA} 
\title{Non-Commutative Inflation}   
\maketitle    
\begin{abstract} We show how a radiation dominated universe subject to space-time
quantization may give rise to inflation as the radiation temperature
exceeds the Planck temperature. We consider dispersion relations with a 
maximal momentum (i.e. a minimum Compton wavelength, or quantum of space), 
noting that some of these lead to a trans-Planckian branch where
energy increases with {\it decreasing} momenta. This feature translates into negative
radiation pressure and, in well-defined circumstances, 
into an inflationary equation of state.
We thus realize the inflationary scenario without the aid of an inflaton field.
As the radiation cools down below the Planck temperature, inflation gracefully exits
into a standard Big Bang universe, dispensing with a period of reheating.
Thermal fluctuations in the radiation bath will in this case generate curvature 
fluctuations on cosmological scales whose amplitude and spectrum can be 
tuned to agree with observations. 
\end{abstract}   
 ]

\renewcommand{\thefootnote}{\arabic{footnote}} 
\setcounter{footnote}{0} 

%\section{Introduction}

In spite of the success of the inflationary Universe scenario 
\cite{Guth:1981zm} in
solving some of the mysteries of standard cosmology and of
providing a mechanism which explains the origin of density
fluctuations on cosmological scales, a mechanism which to date
has passed all of the observational challenges, we still do
not have a convincing realization of inflation based on
fundamental physics. Moreover, the usual realizations of
inflation based on weakly coupled scalar matter fields (see
e.g. \cite{Lindebook} for a comprehensive review) are
plagued by important conceptual problems \cite{Brandenberger:1999sw}. Thus,
it is of great interest to explore possible realizations of
inflation based on new fundamental physics. In particular, since
inflation may occur at energy scales close to the Planck scale,
it is of interest to consider the implications of the recent
developments in our understanding of physics at the Planck
scale for inflation.

Space-time non-commutativity is one of the key new ideas which
follows from recent developments in string and matrix theory
\cite{Banks:1997vh}. It is thus of great interest to explore the
compatibility of non-commutative space-time structure with
inflation (see \cite{Gu:2001ny,Alexander:2001ck} for ideas on how to solve some
of the problems of standard cosmology without inflation in
non-commutative geometry). Non-commutativity, and 
space-time quantization, in general lead to deformed dispersion relations
(see e.g. \cite{Kempf:2001ac}).
It has  been shown \cite{Chu:2000ww,Easther:2001fi,Kempf:2001fa} 
(see also \cite{Brandenberger:2001wr,Martin:2001xs}) that
this can have important
consequences for the predictions of inflation.
These authors demonstrated that the short distance cutoff given by
modifying the usual commutation relations
\be [x,p]=i\hbar(1+\beta p^{2}) \label{com} \ee
changes the perturbation spectrum due to quantum fluctuations.  Implicit
in this approach is the necessity of an inflaton field generating a
de Sitter phase.  
%Our approach is similar, however, we will not postulate
%an inflaton field, nor modified commutation relations.  Instead we will
%incorporate the thermal behavior of modified dispersion relations due to
%space-time non-commutativity.  This will lead to a modification in the
%thermal fluctuations instead of quantum fluctuations.  It may be that
%these two approaches share some similarities and it is the subject of
%future work to address this.
  
In this Letter, we
go one step further and identify dispersion relations for ordinary
radiation which lead to inflation, without the need to introduce a
new fundamental scalar field. Thermal fluctuations then replace
quantum fluctuations as the seeds of cosmic structure.

The dispersion relations derived from the non-commutative structure
of space-time have the property
that there is a maximum momentum (corresponding to a minimum
Compton wavelength or quantum of space). Typically all trans-Planckian
energies get mapped into this maximal momentum. However it is 
also possible to write down deformations for which trans-Planckian
energies get mapped into all momenta smaller than this
maximal momentum. In the latter case for a given momentum there
are two energy levels, one sub-Planckian the other trans-Planckian.
Along the trans-Planckian branch as one decreases the momentum of a particle
its energy increases. 

This unusual feature implies that as we expand a box with radiation 
thermally excited into the trans-Planckian branch, and thereby stretch
the wavelength of all particles and decrease their momenta, 
their energies actually increase. Increased bulk energy as a result
of expansion is the hallmark of negative pressure. We follow the
thermodynamical calculation in detail, with a mixture of analytical
(as developed in \cite{Alexander:2001ck}) and numerical methods, to show
that it is possible to generate an inflationary high
energy equation of state for thermalized radiation subject to 
space-time non-commutativity. We identify a class of dispersion relations
for which this occurs. We also find dispersion relations for which
the equation of state corresponds to ``phantom'' matter \cite{Caldwell:1999ew}.

We start by recalling that non-commutativity leads to deformed dispersion relations,
but whereas space-space non-commutativity must introduce
anisotropic deformations, space-time non-commutativity
preserves isotropy. Hence in the latter case, for massless particles,
the dispersion relations may be written:
\be\label{disp}
E^2 - p^2 c^2 f^2 =0
\ee
where $c$ is a constant reference speed, identified with the 
low-energy speed of light. 
We explore dispersion relations of the form:
\be\label{fdisp}
f=1+(\lambda E)^\alpha \, .
\ee 
The case $\alpha=1$ was proposed in \cite{Amelino-Camelia:2000mn}
and its implications considered in \cite{Alexander:2001ck}, and leads to 
a density dependent equation of state $w = p / \rho$ ($p$ and
$\rho$ denoting pressure and energy density, respectively) which diverges 
like $\log(\lambda\rho)$. We also recall that for this model
the color temperature (i.e., the peak of the thermal spectrum) saturates at 
$T_c\approx 1/\lambda$. In the case $\alpha \neq 1$,
the high energy equation of state $w (\rho\rightarrow\infty$)
turns out to be a constant, and this leads to much simpler cosmological 
scenarios. Depending on $\alpha$ we obtain a realization of
varying speed of light
(VSL: \cite{Moffat:1993ud,Albrecht:1999ir}), inflation, or 
phantom matter \cite{Caldwell:1999ew}. We prove this feature by following
the thermodynamical derivations described in \cite{Alexander:2001ck}.

\begin{figure}
\begin{center}
%\leavevmode
\hspace*{-2.1cm}
\epsfxsize=7.5cm
\epsffile{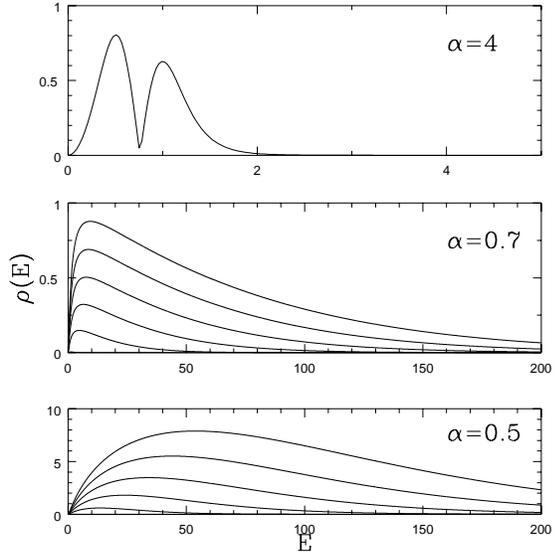}
\end{center}
%\centerline{\psfig{file=spectranew.ps,width=8 cm,angle=0}}
\caption{The thermal spectrum for different temperatures for
 $\alpha=0.5, 0.7, 4$. For $\alpha=0.5$, the peak of the spectrum
scales like the temperature. For $\alpha=0.7$, although the peak
does not vary much with the temperature, the tail of the spectrum
extends to regions which scale with the temperature. For $\alpha=4$
we obtain a double peaked spectrum whose amplitude scales like $T$ (we have
divided all amplitudes by $T$ for convenience), so that
the shape does not depend on temperature.}
\label{spec1}
\end{figure}

As shown in \cite{Alexander:2001ck} the deformed thermal spectrum is given by
\be
\rho(E)={1\over \pi^2\hbar ^3 c^3}{E^3\over e^{\beta E}-1}
{1\over f^3}{\left|
1-{f'E\over f}\right|}\, .
\ee
(note the modulus in the last factor, to be taken whenever the Jacobian of
the transformation $dE/dp$ is not positive definite). This leads to:
\be
\rho(E)={1\over \pi^2\hbar ^3 c^3}{E^3\over e^{\beta E}-1}
{|1+(1-\alpha)(\lambda E)^\alpha|\over (1+(\lambda E)^\alpha)^4}
\ee
We see that the peak
of $\rho(E)$ scales like $T$ for $\alpha<2/3$, as illustrated
in Fig.~\ref{spec1}. For $\alpha>2/3$ the peak saturates
at $E=1/\lambda$, but there is a wide tail up to
$E=T$ for values in the range $2/3<\alpha <1$ (see
Fig.~\ref{spec1}.) For $\alpha>1$ the spectrum becomes double
peaked, with  peaks located 
at $\lambda E \sim 1$ for $\lambda T\gg 1$. The shape of the spectrum
becomes temperature independent since $\rho(E)$ acquires the form
of a temperature independent function of energy  multiplied
by $T$ (see Fig.~\ref{spec1}.) Since the ambient speed of light is given by
$c=dE/dp=c(E)=c(T_{peak})$ we see that
only models with $\alpha<1$ can be implemented
as VSL models. For $\alpha>1$ hotter radiation
means more photons with the same maximal energy, and hence with the same
speed. Only for $\alpha<1$ does hotter radiation mean more energetic and
faster photons, opening doors to VSL.

\begin{figure}
\begin{center}
\leavevmode
\hspace*{-2.1cm}
\epsfxsize=7.5cm
\epsffile{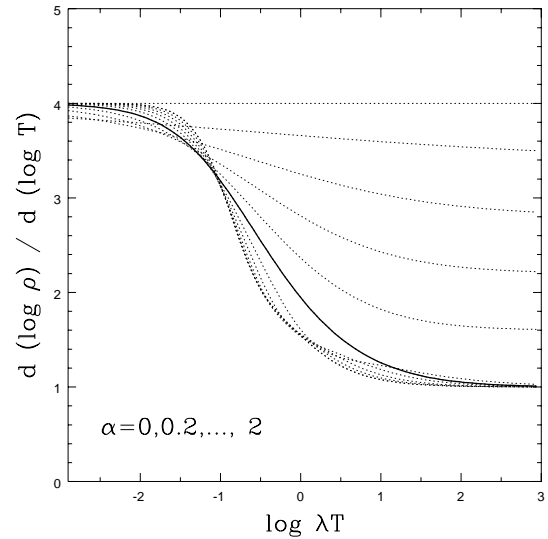}
%\centerline{\psfig{file=stepnew.ps,width=8 cm,angle=0}}
\end{center}
\caption{The high-energy Stephan-Boltzmann
law $\rho\propto T^\gamma$, for different
values of $\alpha$ and different temperatures. The thick line
denotes the case $\alpha=1$ studied in a previous work. We see that
in all cases hotter radiation translates into denser radiation, although
in general $1<\gamma<4$.}
\label{step}
\end{figure}

Next we examine 
whether or not denser radiation means hotter radiation. To answer this question
we integrate $\rho(E)$ to obtain a high-temperature Stephan-Boltzmann
law relating $\rho$ and $T$. We find a power-law of the form
$\rho\propto T^\gamma$, with an asymptotic value for $\gamma$ which varies
from 4 (for $\alpha=0$) to 1 (for all $\alpha\ge 1$). The transition
from low to high temperature behaviour for different values of $\alpha$
is plotted in 
Fig.~\ref{step}. The conclusion is that in all cases denser radiation
corresponds to hotter radiation. 

\begin{figure}
\begin{center}
\leavevmode
\hspace*{-2.1cm}
\epsfxsize=7.5cm
\epsffile{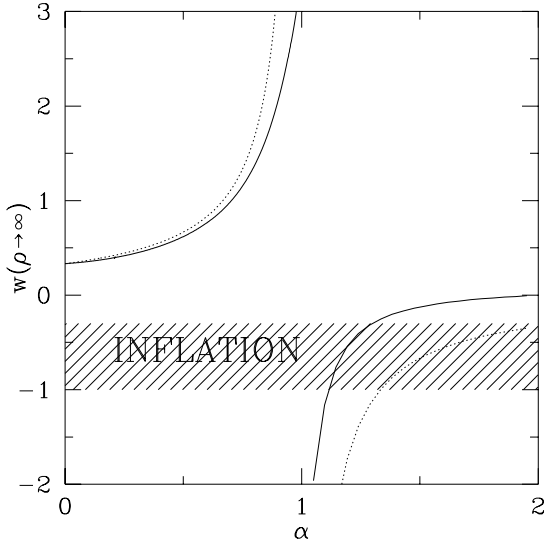}
\end{center}
%\centerline{\psfig{file=wnew.ps,width=8 cm,angle=0}}
\caption{The high energy equation of state for different
values of $\alpha$. We also plotted (dashed line) the approximation
mentioned in the text. The shaded region delimits the values 
of $\alpha$ for which one may have inflationary expansion.}
\label{w}
\end{figure}

Finally the equation of state follows from (see \cite{Alexander:2001ck}):
\be\label{press}
p={1\over 3}\int {\rho(E) dE\over 1-{f'E\over f}}
\ee
Given that the denominator of the integrand is a constant at low and high
energies, we may expect that the high energy equation of state is
a constant approximated by
\be\label{wapprox}
w(\rho\rightarrow\infty)\approx{1\over 3(1-\alpha)}
\ee
Of course, this formula assumes that the peak of $\rho(E)$ is located
at super Planckian energies, where the denominator assumes its
high energy constant value. This does not always happen (e.g.
for $\alpha\ge1$), so a numerical integration of (\ref{press}) is
necessary. We present the result in Fig.~\ref{w}, where we also
plot the approximation (\ref{wapprox}).

We see that in the regime where VSL may be realized ($\alpha<1$)
we do not have inflation, but for $\alpha>1$ we have negative 
pressures. Numbers $1<\alpha_1<\alpha_2$ can be found such that 
for $1<\alpha<\alpha_1$ non-commutative radiation at $T\gg 1/\lambda$
behaves like phantom matter ($w<-1$).  For $\alpha_1<\alpha<\alpha_2$ we have
standard inflationary expansion, with $-1<w<-1/3$, for temperatures
$T\gg 1/\lambda$.

Hence when $\alpha_1<\alpha<\alpha_2$, we have a scenario in which 
the Universe is always filled with radiation, but in the Planck epoch 
(or more precisely when $T\gg 1/\lambda$) radiation 
drives power-law inflation. As a result of expansion this
inflationary radiation cools down, since $w<-1$ implies that
$\rho$ drops with expansion, and $\rho\propto T$ implies that $T$ 
decreases too. 
When the radiation temperature drops below the Planck temperature 
its equation of state reverts to that of normal radiation.
 Then, the 
Universe enters a standard radiation dominated epoch. Our inflationary
scenario does not have
a graceful exit problem, and we have no need of a reheating period.

The critical case $\alpha=\alpha_1$, however, does not benefit
from a graceful exit.
It drives exponential inflation, but the radiation equation of
state is that of a cosmological constant ($w=-1$). As a result
$\rho$ and $T$ stay constant, and the Universe never exits
the de Sitter phase to enter a radiation dominated phase.

The cases $1<\alpha<\alpha_1$ are more complex and will be examined
further elsewhere. For these models
there is a critical $\rho_c$ such that $w(\rho_c)=-1$; for $\rho<\rho_c$
we have $w>-1$ and for $\rho>\rho_c$ we have $w<-1$. If the Universe
starts off trans-Planckian ($\rho>\rho_c$) and expanding, we have 
$a\propto (-t)^{2\over 3(1+w)}$, that is hyper-inflation. However as the
Universe expands it gets denser and hotter (since $w<-1$, and $\rho\propto T$), 
eventually reaching infinite density at $t=0$: perhaps this possibility
may be used to realize the Pre Big Bang scenario \cite{Gasperini:1993em}.
The only regular universe within this case has an infinite de Sitter past 
with $\rho=\rho_c$. However in this model de Sitter space is
unstable. Any small nudge and it either plunges into
an eternal Planck epoch with $\rho>\rho_c$ (with a possible Pre Big Bang
exit) or it decays into a standard radiation epoch. 
%This may be an interesting realization of inflation,
%even though it does require a bit of a fifty fifty anthropic
%argument.

To obtain a heuristic explanation for the origin of negative pressures 
for $\alpha>1$, note that the pressure may be inferred from the change in
the energy inside a box when its size is increased:
\be
p=\sum_s n_s {\left(-\partial E_s\over \partial V\right)}
\ee
where $s$ labels states, $n_s$ their occupation numbers, and
$V=L^3$ the volume of a box of side $L$. The momenta are given
by ${\bf p}={2\pi\hbar\over L}{\bf n}$, where ${\bf n}$ is a triplet
of quantum numbers indexing the state $s$. Hence as the volume increases
the momenta of all states decreases, since their Compton wavelengths
are stretched proportionally to $L$. Usually this translates into a decrease in the energy:
hence the positive pressure of a gas. However for a high temperature
gas living in non-commutative space with $\alpha>1$ the
energy of the dominant branch of the dispersion relation (the higher
energy branch) is a decreasing function of the momentum. Hence a larger
box is reflected in longer Compton wavelengths for these states, and
consequently smaller momenta, but now this implies a larger energy.
Thus, by expanding a box of non-commutative radiation, the system
gains energy, which corresponds to negative pressure.

What is the meaning of the parameters $\alpha$ and $\lambda$? 
Using Equations  (\ref{disp}) and (\ref{fdisp}) we can derive
\be\label{pofe}
p={E \over {c(1+(\lambda E)^\alpha)}}
\ee
from which we see that for $\alpha=1$ there is a maximum allowed momentum
$p_{max}=1/(c\lambda)$.  Its corresponding Compton wavelength is therefore
the minimum length that can be physically probed, 
corresponding to the quantum of space.
Hence $\lambda$ is the parameter determining
the size of the quantum of space.
As $p\rightarrow p_{max}$, the energies $E(p)$ span all possible 
super Planckian energies all the way up to infinity, if $\alpha=1$. 
This changes dramatically if $\alpha>1$: then Equation (\ref{pofe}) also 
shows that there is a maximum momentum; however in 
this case super Planckian energies do not all get mapped into this
momentum - rather we find that for all allowed momenta $p<p_{max}$ 
there are two energy levels, one sub-Planckian the other super-Planckian. 
The function $p(E)$ acquires two branches, along one of which $p$
decreases with $E$. Finally
the case $\alpha<1$ does not contain a sharp maximum
momentum - merely a suppression of variation in momenta with energy
for momenta above a given threshold.

%\section{Cosmological fluctuations}

In our inflationary Universe scenario, it is thermal fluctuations which
are responsible for generating the curvature fluctuations which develop
into the observed perturbations on cosmological scales. A simple way
to estimate the resulting spectrum (see \cite{paper2} for a full
analysis) is to assume fractional thermal density fluctuations of order
unity on the thermal wavelength scale $T^{-1}$. Random superposition
of these fluctuations leads to fractional mass fluctuations on Hubble
radius scale $H^{-1}$ measured at the time $t_i(k)$ that a particular 
wavenumber $k$ crosses the Hubble scale during inflation 
\begin{equation} \label{ampl1}
{{\delta M} \over M}(t_i(k)) = A ({H \over T})^{3/2} \,
\end{equation}
where $A$ is a positive constant smaller than $1$. It is convenient to
express this result in terms of the physical scales $\lambda$ and
$m_{pl}$, and the number $N(k)$ of Hubble times between $t_i(k)$ and
the end of inflation (which roughly occurs when $T = \lambda^{-1}$). 
Application of the Friedmann equations yields
\begin{equation} \label{ampl2}
{{\delta M} \over M}(t_i(k)) = A (\lambda m_{pl})^{-3/2} e^{-3N/(2p)} \, ,
\end{equation}
where $p$ is the power with which the scale factor $a(t)$ increases
during the period of power law inflation.

In order to relate (\ref{ampl1}) with the fractional mass fluctuations
when the scale re-enters the Hubble radius at time $t_f(k)$, we make
use of the fact that fractional density fluctuations increase between
$t_i(k)$ and $t_f(k)$ by a factor given by the ratio of $1 + w$ at the
respective times 
\cite{Bardeen:1980kt,Bardeen:1983qw,Brandenberger:1984tg,Lyth:1985gv}. 
This factor is $2p$. In order to obtain a spectral
slope consistent with the COBE data, the power $p$ has to be sufficiently
large. In this case, requiring that the amplitude of the fluctuations
agree with the data requires $\lambda^{-1}$ to be a couple of orders of
magnitude smaller than $m_{pl}$, which from the point of view of
string theory is not unreasonable.
 
%which always generates hyper-inflation. For $w=-7/3$ we realize exact 
%scale factordualtiy, as in the Veneziano scenario (find numerically

%\section{Conclusions}

In summary, non-commutative space-time geometry leads to modified
dispersion relations. We have identified a class of dispersion relations 
which change the 
high-temperature equation of state of thermal relativistic matter 
into that of inflationary matter. In this scenario inflation
does not require a different type of matter - standard 
radiation suitably heated up will behave like the proverbial inflaton
field. As inflationary expansion proceeds, the radiation
cools down until its equation of state reverts to that of
ordinary radiation and consequently the Universe enters
the standard Hot Big Bang phase. 
Thermal fluctuations in the radiation bath will in this case generate the
necessary density fluctuations to explain the structure of the universe.
Their amplitude and spectrum can be 
tuned to agree with observations. 

%\vskip0.5cm

{\bf{Acknowledgments}}
We would like to thank
W. Unruh for interesting discussions. This work was supported in
part (R.B.) by the US Department of Energy under Contract DE-FG02-91ER40688,
Task A.

\end{document}